# Time Delay in Electron-C$_{60}$ Elastic Scattering in a Dirac Bubble Potential Model

M. Ya. Amusia[1,2] and A. S. Baltenkov[3]

[1]*Racah Institute of Physics, the Hebrew University, Jerusalem, 91904 Israel*
[2]*Ioffe Physical-Technical Institute, St. Petersburg, 194021 Russia*
[3]*Arifov Institute of Ion-Plasma and Laser Technologies*, *Tashkent, 100125 Uzbekistan*

**Abstract**
Within the framework of a Dirac bubble potential model for the C$_{60}$ fullerene shell, we calculated the time delay in slow-electron elastic scattering by C$_{60}$. It appeared that the time of transmission of an electron wave packet through the Dirac bubble potential sphere that simulates a real potential of the C$_{60}$ cage exceeds by more than an order of magnitude the transmission time via a single atomic core. Resonances in the time delays are due to the temporary trapping of electron into quasi-bound states before it leaves the interaction region.

## 1. Introduction

The size of the fullerene C$_{60}$ sphere is much bigger than the size of a single atom. Therefore, the time of an electron wave transmission through the C$_{60}$ cage can significantly exceed the transmission time of an electron wave packet via the atomic potential well. The question "How much time does it take for a particle to tunnel through a potential barrier?" was for the first time formulated almost 90 years ago [1]. The pioneer contributions to this problem came from L. Eisenbud [2], E. Wigner [3] and F. Smith [4], who introduced the time delay (so-called EWS time delay) as a quantum dynamical observable for resonance particle potential *elastic scattering*. This intrinsic time delay is a well-defined characteristic only for short-range interaction of particle wave packet with a target.

Only recently, however, the experimental investigation of the time picture of atomic processes became possible using attosecond laser pulses that opened up the possibility of experimental investigation of time delay in *photoionization processes* [5, 6]. In this case the time-shift relative to XUV pulse "corresponds to a half-scattering process, in which only in the exit channel a matter wave packet resides in the continuum" [6]. In photoionization of usually considered neutral objects, the interaction between the outgoing photoelectron and the residual positive ion is Coulombic. Therefore, the electron wave packet never converges to a free packet and the intrinsic time delay is not a well-defined characteristic of the process. Nevertheless, one can extend the very concept of EWS-time delay to this case also, by separating the long-range electron-ion interaction having the universal character, from the short-range one that is different for different residual ions. Investigation of time delays in *atomic photoionization* is now a rapidly developing field of research (see for example calculations in [5-9] and references therein).

Attosecond spectroscopy of *molecular systems* provides a remarkable possibility of insight into the process of the outgoing wave packet formation within the molecule, where the multi scattering of the electron wave by the atomic constituents is very important. Of special interest are spherically symmetrical fullerenes C$_{60}$ and so-called endohedral atoms A@C$_{60}$. Papers [5, 6] studied the time-resolved photoemission from the central A-atom of molecule A@C$_{60}$, during which the outgoing photoelectron moving in a long-range Coulomb potential crosses the potential of C$_{60}$.

In the present paper, we investigate the partial EWS-time delay in electron elastic scattering by a short-range potential of the fullerene shell, treated in the Dirac bubble potential model of $C_{60}$, initially suggested in [10]. The same times are useful in consideration of photodetachment processes of negative ions of these molecules. Paper [10] represents the potential energy $U(r)$ of the $C_{60}$ shell as a Dirac delta function in the radial coordinate $r$, namely $U(r) = -U_0 \delta(r-R)$, where $R$ is the fullerene radius[*]. Such a model represents the wave functions of electrons in the bound and continuum states as combinations of spherical Bessel functions. Therefore, the expressions for phase shifts and partial time delays in this model are analytical, which is important for simulating the temporal picture of photodetachment of $C_{60}$ negative ion or the resonance electron $e$-$C_{60}$ elastic scattering.

The model [10] permitted to investigate a whole variety of processes with $C_{60}$ participation (see [11] and reviews [12, 13] and references therein) has been investigated. The negative ion $C_{60}^-$ was the initial object studied within the framework of this model. A characteristic feature of $C_{60}^-$ is a relatively low (in atomic scale) affinity energy ($EA$) of an electron to $C_{60}$. Therefore, in the ground state of $C_{60}^-$ the wavelength of the extra electron is larger than the thickness of the spherical layer formed by C atoms of $C_{60}$. This justifies the application of delta-potential $U(r) = -U_0 \delta(r-R)$ and other phenomenological potentials dependent on $r$ only for description of $C_{60}^-$ and collisions of slow electrons with $C_{60}$.

## 2. Time delay in electron-$C_{60}$ elastic scattering

The calculations in paper [11] showed that the cross sections of elastic scattering of slow electrons by $C_{60}$ have a rich resonance structure. As known, the following formula determines the partial EWS-time delay [2], with necessary correction by a factor of two (see e.g. [6])

$$\tau_l(E) = 2 \frac{d\delta_l(E)}{dE}. \qquad (1)$$

Here $\delta_l(E)$ is the phase shift of the electron continuum wave function with orbital moment $l$; $E$ is the electron kinetic energy. Close to resonances, where the potential well captures in fact the incident particle and retains it for some time, the value of $\tau_l(E)$ can be very big.

The Dirac bubble potential model gives the following expression (see Eq. (10) in [11]) for for $e$-$C_{60}$ elastic scattering phase shifts

$$\delta_l = \arctan\left[\frac{x j_l^2(x)}{x j_l(x) n_l(x) - 1/R\Delta L}\right]. \qquad (2)$$

Here $j_l(x)$ and $n_l(x)$ are the spherical Bessel functions with the asymptotic behavior

$$j_l(x \to \infty) = \sin(x - \pi l/2)/x; \quad n_l(x \to \infty) = -\cos(x - \pi l/2)/x, \qquad (3)$$

---
[*]In this paper, we employ the atomic system of units.

where the variable is $x = kR$ with $k = \sqrt{2E}$ being the electron momentum in continuum. The parameter $\Delta L$ in formula (2) is a jump of the logarithmic derivative of the electron wave functions at $r = R$ where the potential $U(r) = -U_0\delta(r-R)$ is infinitely negative. At the point $r = R$ the electron wave functions, for both bound and continuum states, are smooth but their derivatives experience a jump. The following expression $\Delta L = -2U_0$ connects this jump with the strength of the potential $U_0$. The experimental data on $EA$ of $C_{60}^-$ provides information on the $\Delta L$ value. If the extra electron in the $C_{60}^-$ ground state is an $s$-state, then $\Delta L$ is defined by the following expression (for details see [11])

$$\Delta L = \Delta L_{1s} = \left[\frac{d}{dr}\ln K_{1/2}(\beta r)\right]_{r=R} - \left[\frac{d}{dr}\ln I_{1/2}(\beta r)\right]_{r=R}. \quad (4)$$

For the extra electron in the $p$-ground state of $C_{60}^-$ instead of (4) we have

$$\Delta L = \Delta L_{2p} = \left[\frac{d}{dr}\ln K_{3/2}(\beta r)\right]_{r=R} - \left[\frac{d}{dr}\ln I_{3/2}(\beta r)\right]_{r=R}. \quad (5)$$

In formulas (4) and (5) the functions $I_\nu(\beta r)$ and $K_\nu(\beta r)$ are the modified spherical Bessel functions exponentially decreasing with the rise of $r$; $\beta$ is connected with $EA$ by relation $\beta = \sqrt{2I}$ where $I = -EA$. The $p$-like bound state could be a ground one only if one neglects the extra electron interaction with the electromagnetic field. In the considered potential $U(r)$, the $s$- state is always located energetically below the $p$-state. Therefore, the extra electron in the $p$-state decays into the lower $s$-state by emitting a dipole photon.

For the given orbital moment $l$ the wave equation with potential $U(r)$ has a bound state solution (and only one) if the condition $-R\Delta L/2 > l+1/2$ is fulfilled [11] According to this inequality, for $R \approx 6.64$ and $I \approx 0.1$ (2.65 eV), there are only $s$, $p$ and $d$ bound states for our potential well [11]. Thus, in the Dirac bubble potential model the two parameters, namely $R$ and $I$ fully define the whole spectrum of electronic states, both discrete and continuous of $C_{60}$. For the above-given values of $R$ and $I$ the jumps of logarithmic derivatives (4) and (5) are $\Delta L_{1s} \approx -0.885$ and $\Delta L_{2p} \approx -0.993$.

For electron energy $E \leq 0.5$, it is sufficient to take into account the first six phases [11]. Since in our potential well there are $s$-, $p$- and $d$- bound states the phase shifts $\delta_l(k)$ with $l<3$ are equal to $\pi$ at $k \to 0$. Phase shifts for $l \geq 3$ in this limit are $\delta_l(0) \to 0$. The figures 1 and 2 present the results of calculations for scattering phases and corresponding EWS time delays. Figure 1 gives $\delta_l(E)$ and $\tau_l(E)$ for $l=0, 1, 2$, while figure 2 presents $\delta_l(E)$ and $\tau_l(E)$ for $l=3, 4, 5$. Time delays in these figures are given in the atomic units equal to 24.2 $as$, where 1 $as$ = $10^{-18}$ s. The scattering phase shift in a short-range potential should follow the Wigner threshold law. In the case of $s$-phase shift, one has $\delta_0(E \to 0) \propto \pi - E^{1/2}$. Therefore, the time delay in this limit goes to infinity $\tau_s(E \to 0) \propto -E^{-1/2}$. For the orbital moments $l>0$ the time delay vanishes at threshold since $\delta_l(E \to 0) \propto \pm E^{l-1/2}$.

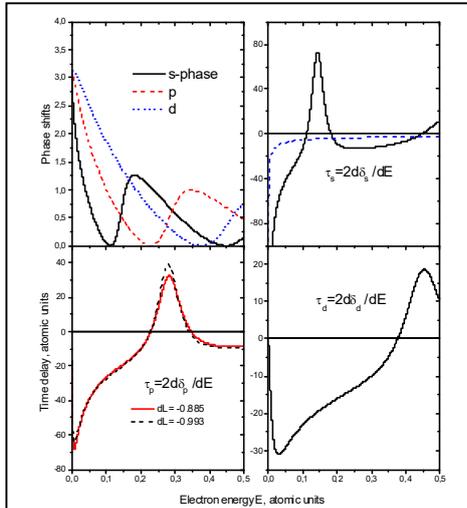

Fig. 1. The phase shifts $\delta_l(E)$ and the EWS-time delays $\tau_l(E)$ for $s$-, $p$-, $d$- partial waves; dashed blue line in the upper right panel is the $s$-time delay for He-atom; dashed black line in the lower left panel are results of calculation $\tau_p(E)$ with the logarithmic derivative of wave functions $\Delta L_{2p} \approx -0.993$.

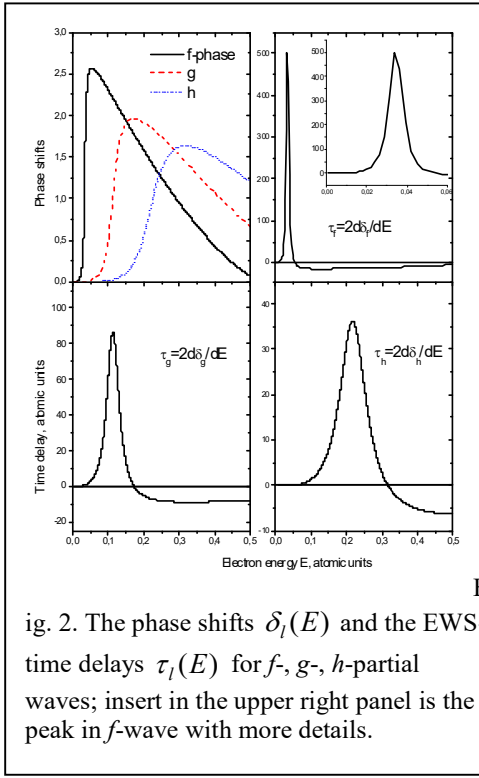

Fig. 2. The phase shifts $\delta_l(E)$ and the EWS-time delays $\tau_l(E)$ for $f$-, $g$-, $h$-partial waves; insert in the upper right panel is the peak in $f$-wave with more details.

The curves in figure 1 have a minimum at small $E$ and reach a positive maximum value with grows of $E$. The height of resonance peaks decreases with the increase in orbital moment $l$. The lower left panel in figure 1 presents $\tau_p$ for two values of $\Delta L$. Since the difference for two respective results is insignificant already for $\tau_p$ and $\tau_d$, we calculated other curves in the figures with $\Delta L = \Delta L_{1s}$. For $\tau_l(E)$ the peak position shifts to higher $E$ with growth of $l$.

The curves in figure 2 have maximums for small electron energy. They result from shape resonances in the partial cross sections of elastic scattering of electrons by $C_{60}$ [11]. Particularly high values reaches the time $\tau_f$ ($\tau_f^{max} \approx 500$) at the electron energy $E \approx 0.035$. The details of the maximum in $\tau_f$ depicts the upper right panel of figure 2. Maxima in $g$- and $h$-waves also exist but they are considerably smaller.

Near maxima, the following expression holds

$$\tau(E) = \tau_{max} \frac{w^2}{4(E-E_r)^2 + w^2}. \qquad (6)$$

Here $E_r$ is the resonance energy, $w$ is the full width of (6) at half maximum (FWHM), $\tau_{max}$ is the maximal value of $\tau(E)$. Table 1 collects the parameters of positive resonance peaks in figures 1 and 2.

**Table 1**. Parameters of the positive resonances

| Partial wave | Resonance energy $E_r$, in $eV$ | Maximal time $\tau_{max}$, in $as$ | FWHM $w$, in $eV$ | $l(l+1)/2R^2$, in $eV$ |
|---|---|---|---|---|
| s | 3.92 | $1.82 \cdot 10^3$ | 1.08 | 0. |
| p | 7.72 | $7.99 \cdot 10^2$ | 2.19 | 0.62 |
| d | 12.39 | $4.53 \cdot 10^2$ | 4.20 | 1.85 |
| f | 0.94 | $1.29 \cdot 10^4$ | 0.21 | 3.70 |
| g | 3.09 | $2.18 \cdot 10^3$ | 1.02 | 6.17 |
| h | 5.90 | $9.05 \cdot 10^2$ | 2.13 | 9.86 |

According to Table 1, the resonance energies $E_r$ of $s$-, $p$- and $d$-partial waves are located above the crest of the centrifugal potential barrier at the point $r=R$, where it is equal to $l(l+1)/2R^2$. The resonance energies for $f$-, $g$- and $h$-partial waves are below the maximal value of this barrier. For these waves the time spent by a particle in the "forbidden" region increases rapidly when the electron energy $E_r$ approaches to the crest of the centrifugal potential barrier. That is in agreement with the calculation results in [14] that demonstrates that particles with much lower and much higher energies (as compared to the potential barrier height) pass through the barrier with negligible delays. While for a particle with an energy, approaching the potential maximum, essential part of its wave function remains under the potential barrier for long period.

It is interesting to compare the time delay for $C_{60}$ bubble potential $\tau_s(E_r)$ in figure 1 with $\tau_s$ for the slow $s$-electron scattering by He atom. The following formula approximates the $s$-phase shift by this atom [15]:

$$\delta_s(k) = \pi - 1.2191k. \qquad (7)$$

The upper right panel of figure 1 gives the EWS-time delay calculated using (7). For $C_{60}$ at $E_r \approx 0.144 \approx 3.92 eV$ the absolute value of the atomic time delay is about 120 $as$. This is 15 times less than the time of electron $s$-wave transmission through the $C_{60}$ cage (see Table 1).

Particle transmission through a potential well is in fact a wave packet transmission of $l$ partial waves with the intrinsic time delay $\tau_l(E)$ for each partial wave. The statistical weight of the $l$-partial wave is the ratio of the partial cross section of electron elastic scattering $\sigma_l(E) = 4\pi(2l+1)\sin^2\delta_l(E)/k^2$ to the total one $\sigma_{tot} = \sum_{i=0}^{\infty} \sigma_l(E)$. Therefore, the averaged time delay by the potential well is

$$\tau_{Aver}(E) = \sum_{l=0}^{\infty} \frac{\sigma_l}{\sigma_{tot}} \tau_l(E). \qquad (8)$$

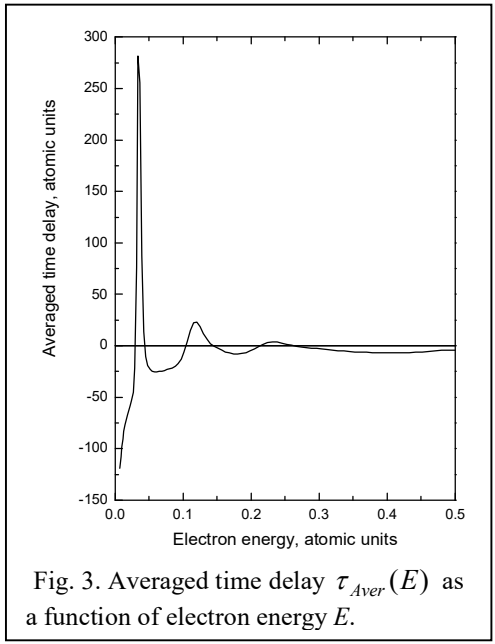

Fig. 3. Averaged time delay $\tau_{Aver}(E)$ as a function of electron energy $E$.

Figure 3 depicts the results of the averaged time calculations. This time reaches a positive maximum within the range of the $f$-partial wave resonance.

3. **Half-scattering process**

Photodetachment of $C_{60}^-$ or endohedral negative ion $A^- @ C_{60}$ are so-called half-scattering processes [5, 6]. In these processes, the interaction between the outgoing photoelectron and the residual molecular core is of short-range and the Coulomb tail is absent. If the bound state of the extra electron in $C_{60}^-$ is an $s$-state, the dipole photodetachment leads to electron emission of only a $p$-wave into the continuum. Therefore, the photoemission time delay $\tau_1(E)$ is only one-half of the given above, i.e. $\tau_1(E) = d\delta_1(E)/dE$. The missing factor of two, as compared to the expression (1) for EWS-time delay, indicates that photoemission is a half-scattering process. According to the lower left panel of figure 1, the photodetachment time delay for $C_{60}^-$ changes within the range from –35 to +20 (from $-8.5 \cdot 10^2$ *as* to $+4.8 \cdot 10^2$ *as*). The times delay by a factor of 20-40 exceeds the typical values for different atomic states having the order of 20 *as* [6].

Paper [16] investigates photodetachment of $A^- @ C_{60}$ that is a negative atomic ion located at the center of the $C_{60}$ shell, in the zero-range potential model for $A^-$. Electron emission into continuum *p*-state accompanies the dipole photodetachment of a free $A^-$. However, for the free $A^-$ ion in the zero-range potential model the *p*-phase shift of the continuum wave function is zero. Therefore, the time delay $\tau_1(E)$ for $A^-$ ion located at the center of the $C_{60}$ sphere is only due to the *p*-wave scattering by the $C_{60}$ bubble potential. Figure 4 illustrates the connection of resonance in the *p*-time delay with resonance in photodetachment cross section. The following expression describes the photodetachment cross section of the free negative $A^-$ ion in the zero-range potential model

$$\sigma_{free} = \frac{16\pi}{3} \alpha \frac{\sqrt{I_{A^-}}(\omega - I_{A^-})^{3/2}}{\omega^3}. \quad (9)$$

Here $\alpha$ is the fine structure constant, $\omega = I_{A^-} + k^2/2$ is the photon energy, $I_{A^-}$ is the detachment energy of a negative ion that we put here equal to $I$=3.4 eV. This choice is good for so-called simple negative ions, e.g. such as $Na^-$. The following formula describes the photodetachment of the $A^-$ ion located at the center of $C_{60}$

$$\sigma_@(\omega) = \sigma_{free}(\omega) D_1^2(k). \quad (10)$$

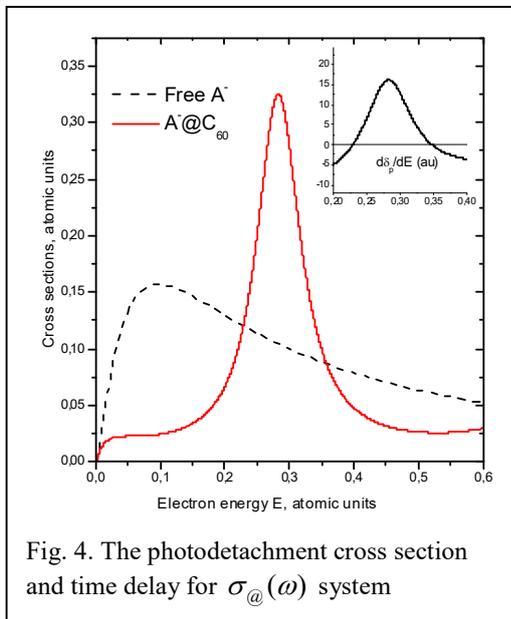

Fig. 4. The photodetachment cross section and time delay for $\sigma_@(\omega)$ system

Here the amplitude factor $D_1(k)$ looks like (see Eq. (7) in [16])

$$D_1(k) = \cos\delta_p(k)\left[1 - \tan\delta_p(k)\frac{n_1(x)}{j_1(x)}\right]. \quad (11)$$

Figure 4 presents the cross-section $\sigma_@(\omega)$ and time delay $\tau_@$ for photoionization of $A^- @ C_{60}$. At the maximum of $\sigma_@(\omega)$, the time delay reaches almost 400 *as*.

Performed in this paper analysis makes it possible to answer one of the key questions of the modern time delay problem (see [6], page 785): "Does it take longer for the electron to escape from the multicenter molecular core than from the one-center atomic core?". Note that the question rhymes with that formulated almost 90 years ago [1]. Summarizing the results of the calculations performed, we conclude that the transmission time of electron wave packet through the Dirac bubble potential sphere (simulating the real short-range potential of the C$_{60}$ cage) exceeds by more than an order of magnitude the transmission time for the one-center atomic core. Resonance electron scattering by short-range potential in this model is due to the transient trapping of an electron in quasi-bound *f*-, *g*- and *h*-states before leaving the interaction region as an outgoing wave packet.

The time delays in photoelectron emission from $C_{60}^-$ or endohedral $A^-@C_{60}$ are also much bigger than the atomic ones. This is an evidence of significant environment-specific contributions of the C$_{60}$ shell to the time shift. Variation of the bubble potential parameters opens up the opportunity to apply this model (along with the Yukawa model [6]) to simulate and explore in natural time scale the electron interaction not only with fullerene C$_{60}$ but also with a whole class of spherical carbon molecules – onion-like fullerenes or single- and multi-walled doped fullerenes [13] and endohedrals [17].

It is of interest to study time delays in elastic scattering processes experimentally. However, one should keep in mind that the duration of a quantum-mechanical process determined by (1) differs from respective value in classical physics, since uncertainty principle puts its limitation upon accuracy of definition of both duration $\tau(E)$ and energy *E*, namely $\Delta\tau \cdot \Delta E > \hbar$ .

Nowadays, attosecond lasers are the main tool in studies of the temporal picture of atomic processes. Therefore, it is essential to keep in mind that the uncertainty in energy introduced by a typical attosecond incoming energy pulse is as big as several hundreds of eV. It means that such a pulse, while applied to an atom or a molecule, engages in a reaction a number of processes simultaneously, thus making it difficult to compare experimental results with calculation data describing energy resolved separated reactions.

**Acknowledgments**
ASB is grateful for the support to the Uzbek Foundation Award OT-Ф2-46.

References
1. L. A. MacColl, Phys. Rev. **40**, 621 (1932).
2. L. E. Eisenbud, *Ph. D. thesis*. Princeton University (1948).
3. E. P. Wigner, Phys. Rev. **98** 145 (1955).
4. F. T. Smith, Phys. Rev. **118** 349 (1960).
5. R. Pazourek, S. Nagele, and J. Burgdörfer, Faraday Discuss. **163** 353 (2013).
6. R. Pazourek, S. Nagele, and J. Burgdörfer, Rev. Mod. Phys. **87** 765 (2015).
7. A. S. Kheifets and I. A. Ivanov, Phys. Rev. Lett. **105**, 233002 (2010).
8. A. S. Kheifets, Phys. Rev. A **87**, 063404 (2013.)
9. P. C. Deshmukh, A. Mandal, S. Saha, A. S. Kheifets, V. K. Dolmatov, and S. T. Manson, Phys. Rev. A **89**, 053424 (2014).
10. L. L. Lohr and S. M. Blinder, Chem. Phys. Letters **198** 100 (1992).
11. M. Ya. Amusia, A. S. Baltenkov, B. G. Krakov, Phys. Letters A **243** 99 (1998).
12. V. K. Dolmatov, A. S. Baltenkov, J. P. Connerade, S. T. Manson, Radiat. Phys. Chem. **70** 417 (2004).
13. V. K. Dolmatov, Advances in Quantum Chemistry **58** 13 (2009).
14. A. Goldberg and H. M. Schey and J. L. Schwartz, Am. J. Phys. **35** 177 (1967).
15. N. F. Mott and H. S. W. Massey, *The Theory of Atomic Collisions* (Clarendon Press, Oxford, 1965).
16. A. S. Baltenkov, Phys. Letters A **254** 203 (1999).
17. M. Ya. Amusia, L. V. Chernysheva, and E. Z. Liverts, Phys. Rev. A **80,** 032503-1-12 (2009).

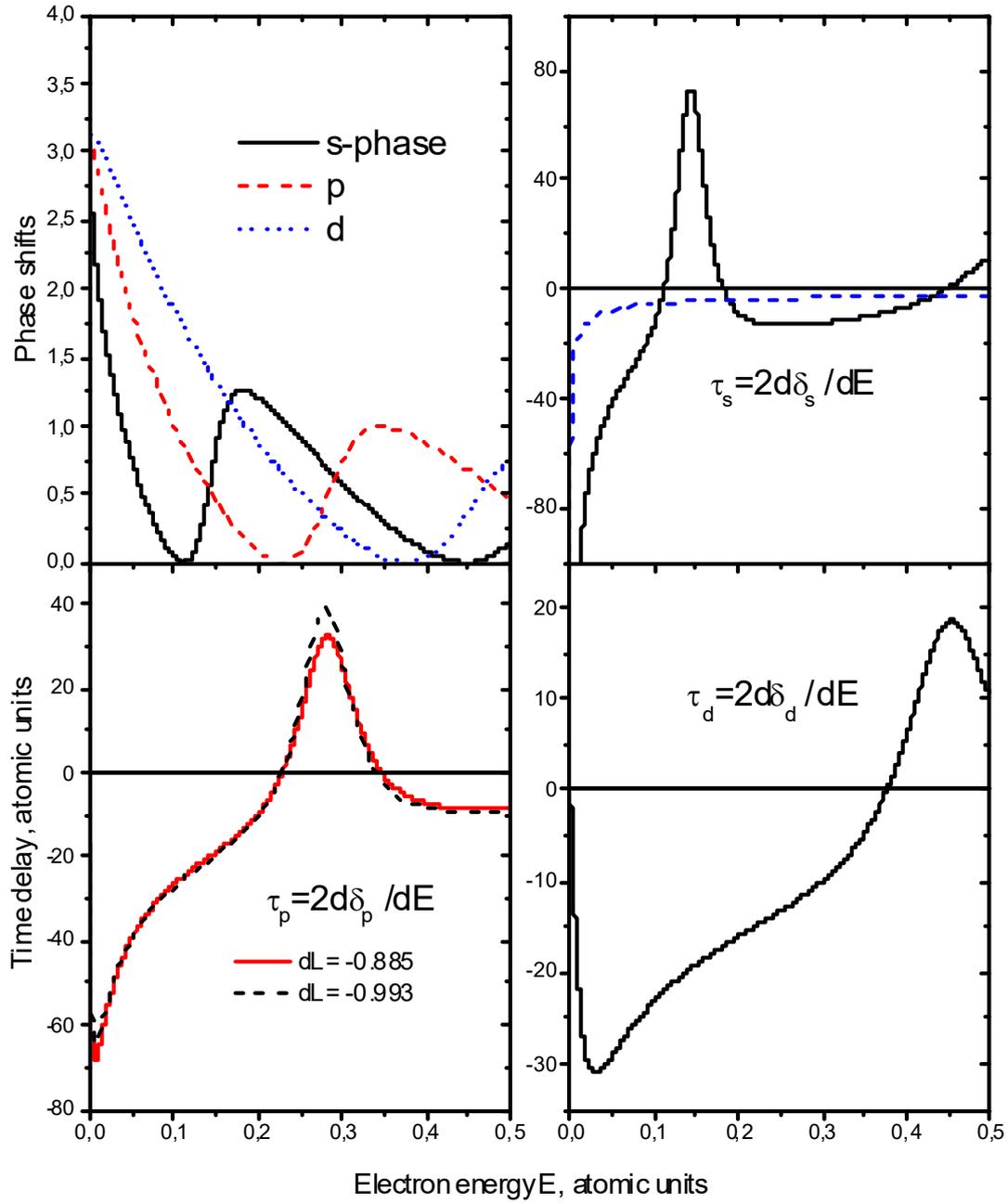

Fig. 1. The phase shifts $\delta_l(E)$ and the EWS-time delays $\tau_l(E)$ for *s*-, *p*-, *d*-partial waves; dashed blue line in the upper right panel is the *s*-time delay for He-atom; dashed black line in the lower left panel are results of calculation $\tau_p(E)$ with the logarithmic derivative of wave functions $\Delta L_{2p} \approx -0.993$.

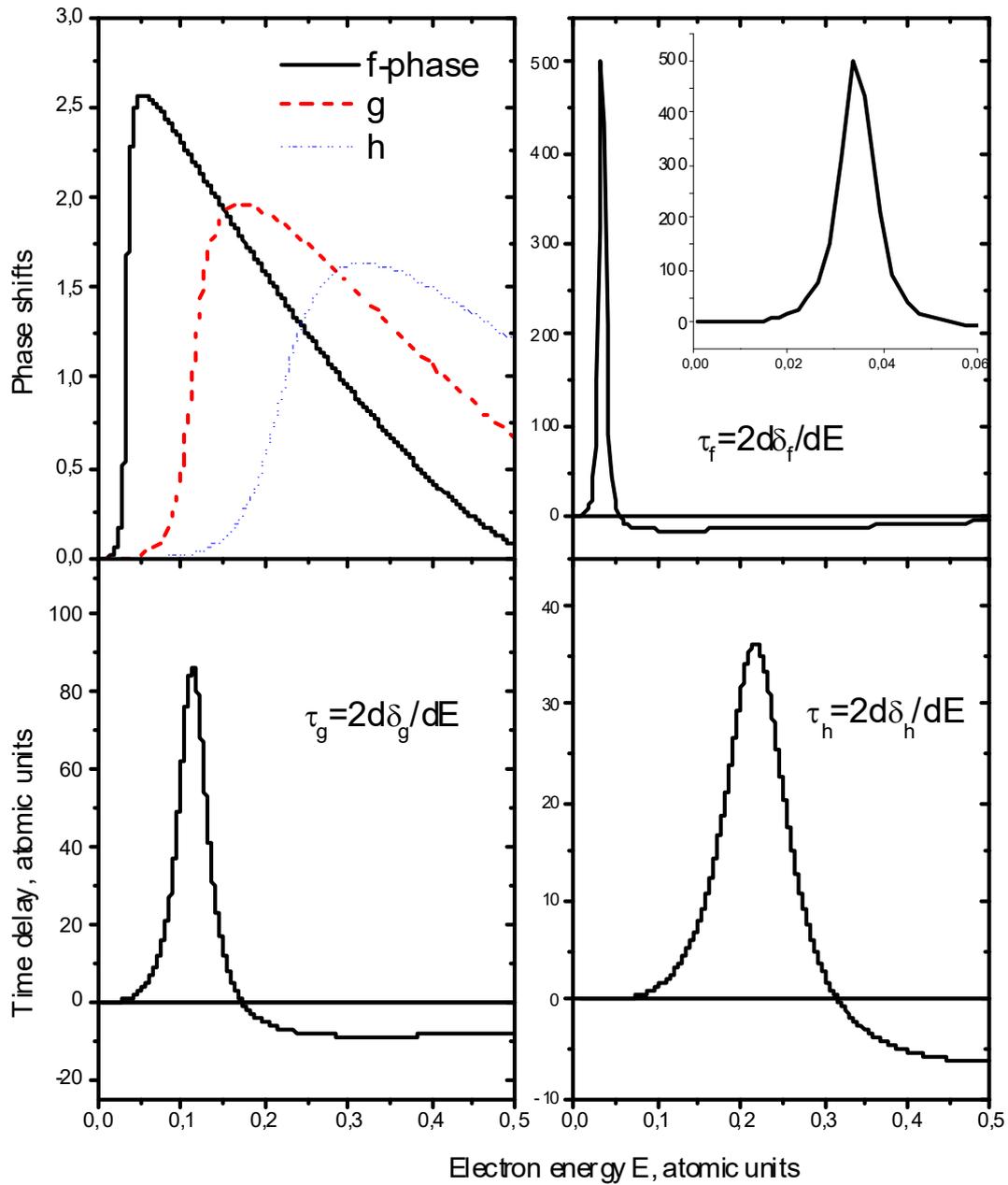

Fig. 2. The phase shifts $\delta_l(E)$ and the EWS-time delays $\tau_l(E)$ for *f*-, *g*-, *h*-partial waves; insert in the upper right panel is the peak in *f*-wave with more details.

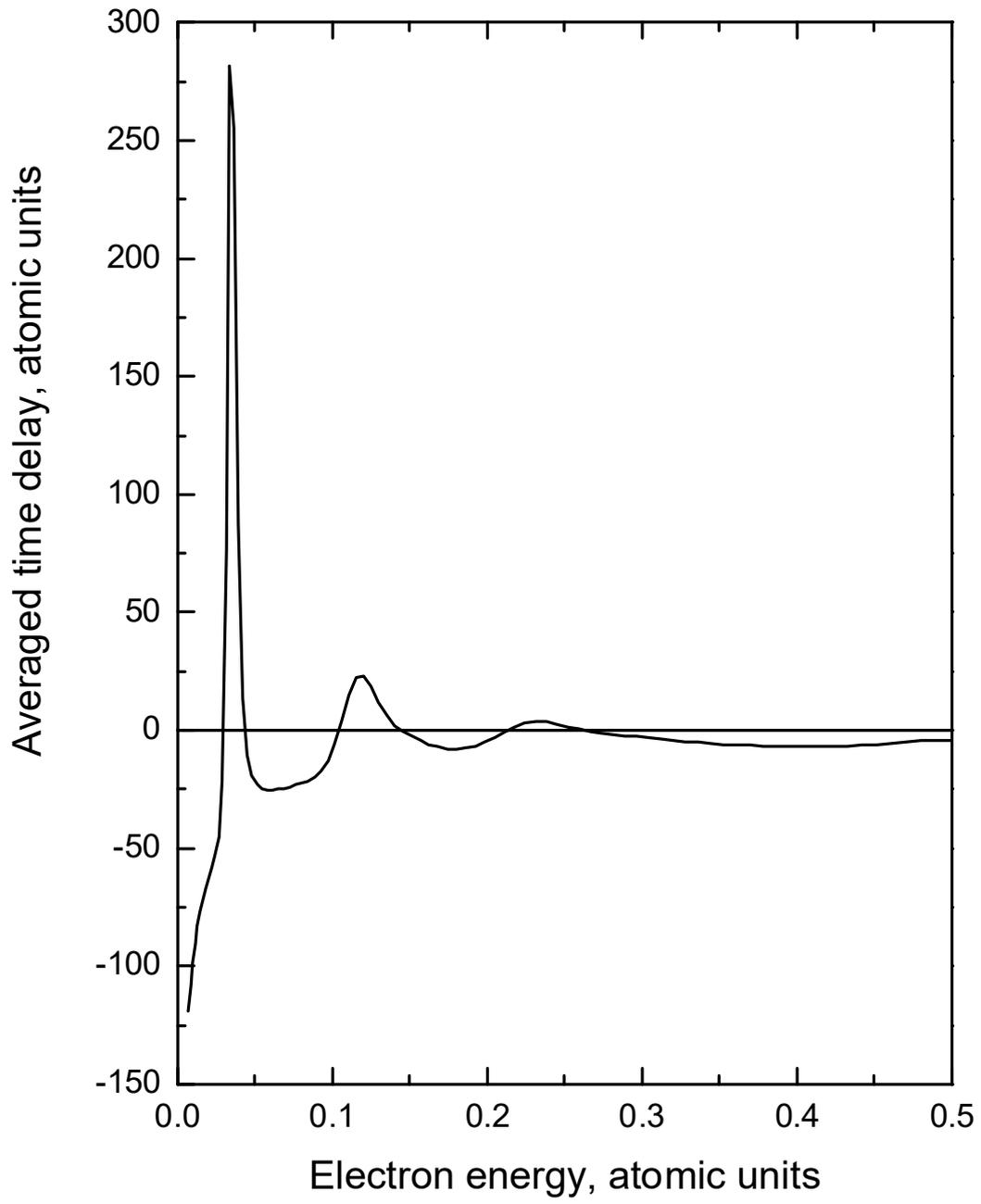

Fig. 3. Averaged time delay $\tau_{Aver}(E)$ as a function of electron energy $E$.

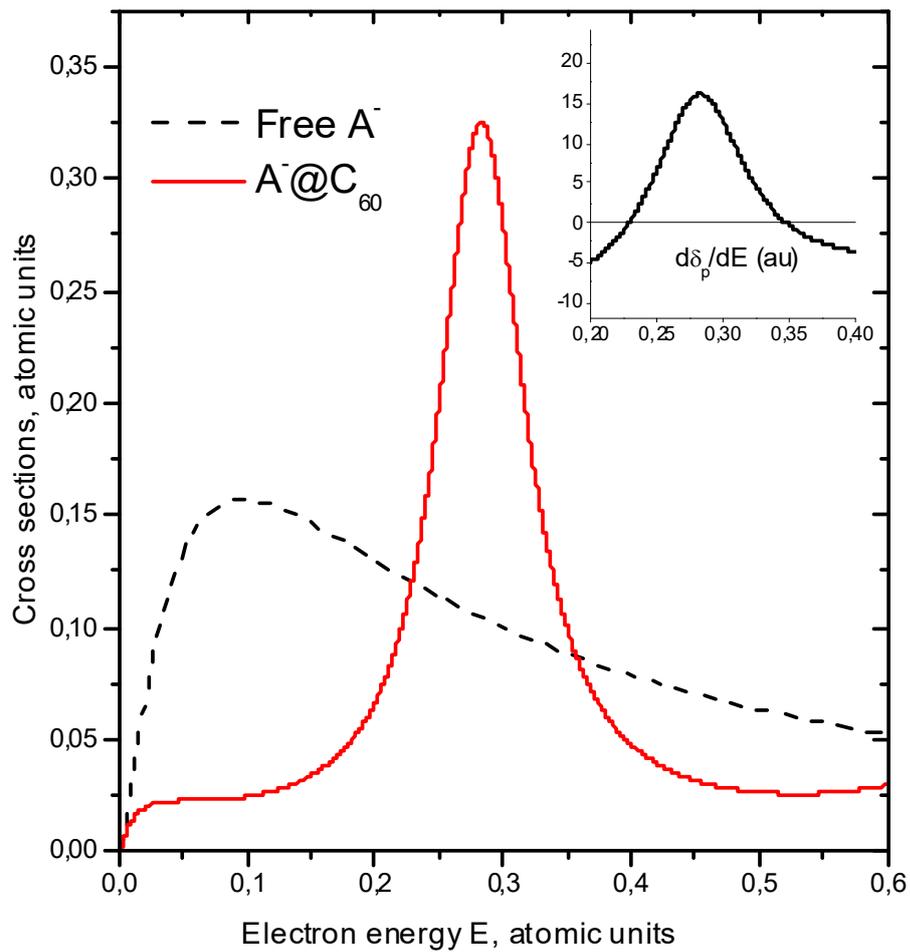

Fig. 4. The photodetachment cross section and time delay for $\sigma_{@}(\omega)$ system